 \documentclass[final,3p,twocolumn]{elsarticle}

\usepackage{amssymb}
\usepackage{graphicx}
\usepackage{dcolumn}
\usepackage{subfigure}
\usepackage[normalem]{ulem}
\usepackage{bm}
\usepackage{fancyhdr}
\usepackage{lastpage}
\usepackage{graphicx, wrapfig, setspace, booktabs}
\usepackage[T1]{fontenc}
\usepackage[font=small, labelfont=bf]{caption}
\usepackage[protrusion=true, expansion=true]{microtype}
\usepackage{hyperref}
\usepackage{url,lipsum}
\usepackage{epstopdf,epsfig}
\usepackage{amsmath,amsfonts,amssymb}
\usepackage{float}
\usepackage{url}
\usepackage{natbib}
\usepackage{todonotes}

\newtheorem{theorem}{Theorem}

\begin{document}

\begin{frontmatter}



\title{Energy self-balance as the physical basis of orbit quantization}


\author[inst1]{Álvaro G. López}
\ead{alvaro.lopez@urjc.es}
\author[inst2]{Rahil N. Valani}


\affiliation[inst1]{organization=  {Nonlinear Dynamics, Chaos and Complex Systems Group.\\Departamento de F\'isica},
            addressline={Universidad Rey Juan Carlos, Tulip\'an s/n}, 
            city={M\'ostoles},
            postcode={28933}, 
            state={Madrid},
            country={Spain}}

\affiliation[inst2]{organization={Rudolf Peierls Centre for Theoretical Physics, Parks Road,
University of Oxford},
            city={Oxford},
            postcode={OX1 3PU},
            country={United Kingdom}}

\begin{abstract}
We show that work done by the non-conservative forces along a stable limit-cycle attractor of a dissipative dynamical system is always equal to zero. Thus, mechanical energy is preserved on average along periodic orbits. This balance between energy gain and energy loss along different phases of the self-sustained oscillation is responsible for the existence of quantized orbits in such systems. Furthermore, we show that the instantaneous preservation of projected phase-space areas along quantized orbits describes the neutral dynamics of the phase, allowing us to derive from this equation the Wilson–Sommerfeld-like quantization condition. We apply our general results to near-Hamiltonian systems, identifying the fixed points of Krylov-Bogoliubov radial equation governing the dynamics of the limit cycles with the zeros of the Melnikov function. Moreover, we relate the instantaneous preservation of the phase-space area along the quantized orbits to the second Krylov-Bogoliubov equation describing the dynamics of the phase. We test the two quantization conditions in the context of hydrodynamic quantum analogs, where a megastable spectra of quantized orbits have recently been discovered. Specifically, we use a generalized pilot-wave model for a walking droplet confined in a harmonic potential, and find a countably infinite set of nested limit cycle attractors representing a classical analog of quantized orbits. We compute the energy spectrum and the eigenfunctions of this self-excited system.
\end{abstract}



\begin{keyword}
Self-oscillation \sep Orbit quantization \sep Hopf bifurcation \sep Hydrodynamic quantum analogs \sep Walking droplets \sep Active matter
\end{keyword}

\end{frontmatter}


\section{Introduction}

Recent studies in hydrodynamic quantum analogs have demonstrated that classical orbit quantization is feasible \cite{Fort_2013,labousse2016b}. In experiments, a millimeter-sized droplet of silicone oil walks on a vibrating bath of the same liquid while bouncing periodically. On each bounce, the droplet imprints a slowly decaying, localized, standing pilot wave on the liquid surface. The droplet in turn interacts with these self-generated waves on subsequent bounces, resulting in a time-delayed self-excited force that propels itself horizontally. Thus, the droplet and its pilot wave coexist inextricably as a wave-particle entity, and the former is active (in the sense of active matter), because it locally extracts energy from the vertical vibration of the bath and converts it into horizontal self-propelled motion.

The hydrodynamic analog of quantization with walking droplets is experimentally realized by confining the wave-particle entity in a two-dimensional harmonic potential, which results in a discrete but \emph{finite} set of stable periodic orbits such as circles, ovals, lemniscates, and trefoils, as the width of the potential is varied~\citep{Perrard2014a,Perrard2014b,Tambasco2016,Tambascoorbit,Labousse_2014,labousseharmonic,Kurianskiharmonic,durey2018,Perrard2018}. The importance of memory for orbit quantization in non-Markovian models of this system is well-established \cite{Labousse_2014}, since the droplet retains path memory from the slowly decaying self-generated waves that sculpt its complex dynamical landscape. A minimal model based on wave interference has recently been developed to explain the origin of quantization \cite{Blits24}. The effective net force that guides the dynamics of the walking droplet can be simplified to two contributions consisting of local instantaneous forces produced by the most recent bounces, and spatio-temporal nonlocal delayed forces coming from past stationary points of the walker's trajectory. The former forces account for the preferred speed of the walker and its wave-induced increased inertia, while the latter are responsible for orbital quantization \cite{Blits24}. However, although these works address essential dynamical aspects and key mechanisms of orbit quantization in two-dimensional walking droplet models, they do not explain how the continuous infinite spectrum of possible energy values is discretized into the few limit-cycle orbits, neither why only a few of them coexist. Since the proposed mechanism of wave interference is also present during the transient dynamics towards the stable periodic orbits, it is not clear why the trajectories should settle on orbits with such precise mechanical (average) energy values, while others are forbidden. Moreover, the possible existence of a full quantized spectrum \cite{LopezValani2025}, i.e. megastability, is not considered. 

In this Letter, we investigate the physical basis of discretization of a continuous energy spectrum to a countably infinite set of stable limit cycles \cite{Sprott2017} from a general viewpoint. We mathematically demonstrate that the balance of energy due to non-conservative forces during self-oscillation \cite{jenkins} is the key thermodynamic constraint producing the orbit quantization \cite{lopezte}. This restriction allows us to connect classical orbit quantization to the original Bohr-Sommerfeld-Wilson model of quantum mechanics \cite{bohr13,Wilson01061915}. To substantiate our analytical findings, we test numerically the main results in a stroboscopic model of a walking droplet with truncated memory and a Bessel function kernel, confined to a one-dimensional harmonic potential, in the low dissipation and low-memory regime.

\section{Main Result}\label{Sec: DS}
Consider the dynamics of a classical particle of mass $m$ moving in $3$D with position $\vec{x}$ and velocity $\dot{\vec{x}}$. The particle is under the influence of an external conservative potential $V(\vec{x})$ and a general non-conservative force given by $- \vec{g}(\epsilon,\vec{x},\dot{\vec{x}})$ that could be a source of both self-excitation and dissipation, with $\epsilon$ a tunable parameter. From Newton's second law, the dynamics of the particle is governed by
\begin{equation}
    m \ddot{\vec{x}}+ \vec{g}(\epsilon,\vec{x},\dot{\vec{x}}) + \vec{\nabla} V(\vec{x}) = 0.
   \label{eq:1}
\end{equation}
If we define the Lyapunov energy function as the energy associated with the conservative part of the equation of motion
\begin{equation}
    H(\vec{x},\dot{\vec{x}})=\frac{1}{2} m \dot{\vec{x}}^{2}+ V(\vec{x}),
   \label{eq:2}
\end{equation}
and we further introduce the linear momentum $\vec{y}=m \dot{\vec{x}}$, this dynamical equation can be written in phase space as
\begin{align}
 \label{eq:3}
 & \dot{\vec{x}}=\vec{\nabla}_{y} H+ \vec{g}_x(\epsilon,\vec{x},\vec{y}),\\ \nonumber
 &\dot{\vec{y}}=-\vec{\nabla}_{x} H+ \vec{g}_y(\epsilon,\vec{x},\vec{y}).
\end{align}
with $\vec{g}_x(\epsilon,\vec{x},\vec{y})=0$ and $\vec{g}_y(\epsilon,\vec{x},\vec{y})=-\vec{g}(\epsilon,\vec{x},\vec{y})$. 
Thus, our system does not conserve energy due to the presence of $\vec{g}(\epsilon,\vec{x},\vec{y})$. The system in \eqref{eq:3} describes a dissipative structure open to the environment, and its dynamics is governed by a \emph{perturbed} Hamilton's canonical equations of motion. Typical motion of Eq.~\eqref{eq:3} results in stable periodic orbits, i.e., limit cycles, quasiperiodic attractors, or even strange chaotic attractors \cite{VALANI2024115253}.

We now prove the following result:

\begin{theorem}
Let the dynamical system in Eq.~\eqref{eq:3} be given in any arbitrary number of dimensions $2N$, with $H$ defined as in Eq.~\eqref{eq:2}, and let $\vec{g}(\epsilon,\vec{x},\dot{\vec{x}})$ be a $C^{1}$ differentiable function of the phase-space variables. Then, any periodic solution of Eq.~\eqref{eq:3} conserves $H$ on average and instantaneously preserves the enclosed phase-space area projections.
\end{theorem}

The proof of the first part of this theorem is self-evident. Since for conservative fields of force the potential energy is preserved along a closed trajectory, the net work performed by the external conservative forces along the periodic limit cycle is zero. Any loss or gain in mechanical energy must then be due to dissipative or self-excited forces. Thus, we can further relate the instantaneous rate of gain or loss of mechanical energy $\dot{H}$ to the work done by the self-excited force. The latter is written as 
\begin{align}
\label{eq:4}
W_{\text{self}}=-\oint \vec{g}(\epsilon,\vec{x},\dot{\vec{x}}) d \vec{x},
\end{align}
yielding the \emph{first quantization condition} in terms of the integral constraint
\begin{align}
\label{eq:4}
\int^{T}_{0} \vec{g}(\epsilon,\vec{x},\dot{\vec{x}}) \dot{\vec{x}} d t = 0.
\end{align}

This means that along a quantized orbit the average power of the self-excited forces always equals zero \cite{poincare1908}. Since the limit cycle orbits obey Eq.~\eqref{eq:3}, by noting that average energy conservation is written as $\langle \dot{H} \rangle$, we can express the last integral as
\begin{align}
 \oint d H=\int^{T}_{0} \left( \vec{\nabla}_{x} H \cdot  \dot{\vec{x}} + \vec{\nabla}_{y} H \cdot  \dot{\vec{y}}\right)d t =  W_{\text{self}},
 \label{eq:5}
\end{align}
where Eq.~\eqref{eq:3} has been used to replace the gradients of $H$ in the last equality of Eq.~\eqref{eq:5}. Thus we conclude that $\langle \dot{H} \rangle = \oint d H/T = 0$ for any limit cycle orbit. 

Moreover, energy is conserved on average if and only if the Melnikov function is equal to zero \cite{sanjuan1998lienard,Davidow2017}. Indeed, substituting Eq.~\eqref{eq:3} in Eq.~\eqref{eq:5} we get 
\begin{align}
\label{eq:6}
M(\epsilon) \equiv \int^{T}_{0} (\vec{\nabla}_{x} H \cdot \vec{g}_x + \vec{\nabla}_{y} H \cdot \vec{g}_y) d t =0,
\end{align}
where we have defined $M(\epsilon)$ as the Melnikov function of the system of Eqs.~\eqref{eq:3}. The Melnikov distance is crucially important, as it is used to determine the existence of homoclinic intersections \cite{guckenheimer2013nonlinear} in dynamical systems, allowing us to generalize our energy self-balance quantization condition to non-integrable chaotic dissipative dynamical systems (see Sec.~5). 

Summarizing, Eq.~\eqref{eq:4} entails that the work done by the self-excited force is equal to zero along the quantized orbit. This, in turn, means that \emph{quantized orbits are the only possible orbits along which the self-forces balance}. As we illustrate below, this energy selection rule restricts the continuous spectra of energy levels to a set of at most countably infinite discrete values, constraining the possible trajectories of the particle. Any other path involves a disturbance of the energy balance, making the trajectories unstable and leading, after transient dynamics, to one of the stable quantized orbits \cite{LopezValani2025}. 

The second part of the proof is not so immediate. Firstly, we recall that the phase-space area projections $(x_k(t),y_k(t))$ along limit cycle orbits define closed curves, where $k$ denotes a particular dimension of our system. Given an initial area projection in the phase space $A_k(0)$, it can be calculated using the integral $A_k(0)= \iint_{\Omega_{0}} d x_k \wedge d y_k$, where $\Omega_0$ is the initial domain. It is well known that the rate of change in area $A_k(t)$ at that particular instant $t$ is given by the integral of the Jacobian trace $J_k$ restricted to the subspace spanned in dimension $k$. Thus the evolution of the phase space area projections are governed by the equations
\begin{align}
\label{eq:7}
 \frac{d A_k}{d t}= \iint_{\Omega} \text{tr}(J_k) d A_k.
\end{align}

In the present case we have that $\text{tr}(J_k)= \partial g_{1 k}/\partial x_k+\partial g_{2k}/\partial y_k=\partial g_k/\partial y_k$, so that by choosing the area enclosed by a limit cycle trajectory, and using Stokes' theorem, we get
\begin{align}
\label{eq:8}
 \frac{d A_k}{d t}= \int^{T}_{0} g_k(\epsilon,\vec{x},\dot{\vec{x}}) \dot{y}_k d t,
\end{align}

For each dimension $k$, Eq.~\eqref{eq:8} must be equal to zero for a limit cycle trajectory, since the enclosed area does not change along the invariant projected curve. In this way, the limit cycle trajectories preserve the projected areas instantaneously ($\dot{A_k}=0$), and therefore their sum is also conserved. Thus, gathering all the dimensions from $k=1,...,N$, the \emph{second quantization condition} can be written as
\begin{align}
\label{eq:9}
\int^{T}_{0} \vec{g}(\epsilon,\vec{x},\dot{\vec{x}}) \dot{\vec{y}} d t=0,
\end{align}
and corresponds to a Sommerfeld-Wilson-Ishiwara quantization rule \citep{sommerfeld1923,Wilson01061915}. This is a consequence of the fact that the field lines of the flow become tangent to the limit cycle. From the point of view of differential geometry, the net flux of the field appearing in Eq.~\eqref{eq:3} across the limit cycle is zero, which implies that the area projections of such a limit cycle never expand or contract beyond their limits.

\section{Krylov-Bogoliubov method}\label{Sec: KB}

In the case of one-dimensional near-Hamiltonian systems \cite{han2009hopf}, we can identify the previous quantization conditions of Eqs.~\eqref{eq:4} and~\eqref{eq:9} with the equations appearing in the Krylov-Bogoliubov averaging method \cite{krylov1950}. This method proposes the following ansatz $$(x(t),y(t))=(r(t) \sin(t+\varphi(t)),r(t) \cos(t+\varphi(t)))$$ for Eq.~\eqref{eq:3} with $g_2(\epsilon,x,y)=\epsilon g(x,y)$ in the limit $\epsilon \ll 1$, yielding the equations
\begin{align}
\dot{r}(t) & = \epsilon g(r(t) \sin \theta(t), r(t) \cos \theta(t)) \cos \theta(t),    
\nonumber \\ 
r(t)\dot{\varphi}(t) & = -\epsilon g(r(t) \sin \theta(t), r(t) \cos \theta(t)) \sin \theta(t), \nonumber
\end{align}
where the variable $\theta(t)=\varphi(t)+t$ has been introduced. The time variable can be averaged over the period $T$, corresponding to $\theta(T)=2\pi$. To obtain separate differential equations for the radius and phase of the limit cycles, we take advantage of the fact that, for small $\epsilon$, a scale separation between the radius and the phase of the limit cycles is allowed \cite{krylov1950}. This yields the two following integrals
\begin{align}
\dot{r}(t) & = \frac{\epsilon}{2 \pi} \int^{2 \pi}_{0}  g(r(t) \sin \theta, r(t) \cos \theta) \cos \theta d \theta, \label{eq:11_1} \\ 
r(t) \dot{\varphi}(t) & = -\frac{\epsilon}{2 \pi} \int^{2 \pi}_{0} g(r(t) \sin \theta, r(t) \cos \theta) \sin \theta d \theta. 
\label{eq:11_2}
\end{align}
The approximation error remains bounded and can be reduced arbitrarily by decreasing $\epsilon$~\citep{krylov1950}. Note that the fixed points of Eqs.~\eqref{eq:11_1} and ~\eqref{eq:11_2} are equivalent to the first and second quantization conditions appearing in Eqs.~\eqref{eq:4} and \eqref{eq:9}, respectively. In this manner, we uncover the relation between general quantization rules and averaging methods in nonlinear dynamics. The first quantization condition for weakly perturbed systems is mathematically equivalent to computing the fixed points of the radii $r_n$ of the stable limit cycles. The fixed points of the phase correspond to the second equation of Krylov-Bogoliubov, which is related to the preservation of the area enclosed by the limit cycles under the evolution of the flow.

Importantly, this occurs even for $\epsilon$ arbitrarily small, suggesting that mechanical systems based on conservative Hamiltonian systems with continuous spectra of energies should be considered more the exception than the rule. Furthermore, the Krylov-Bogoliubov method can be used to derive a Sommerfeld-Wilson-Ishiwara quantization rule, by computing the action $J_n$ representing the phase space area along the quantized orbits
\begin{align}
\label{eq:12}
J_n=\oint y d x = r^2_n \int^{2 \pi}_0 \cos^2\theta d\theta = \pi r^2_n,
\end{align}
where $r_n$ is the radius of the $n$-th quantized orbit, obtained as a solution to Eq.~\eqref{eq:11_1}. Thus, according to the Krylov-Bogoliubov method, $n$ can be considered an adiabatic invariant, as originally suggested by Sommerfeld \citep{sommerfeld1923}. 

This phase space volume preservation is equivalent to stating that the Lyapunov exponent along the limit-cycle trajectory is zero. The fact that the phase is neutral ($\dot{\varphi}=0$) is a well-known property of self-oscillators \cite{jenkins}. It is crucial to enable the mechanism of phase synchronization of oscillations \cite{pikovsky1985universal}, suggesting a physical mechanism of entangled states and quantum-like correlations \cite{papatryfonos2024static}. This contrasts to resonant periodically driven nonlinear oscillators that are not self-sustained and which have their ``intrinsic'' phase locked to the external perturbation for any possible initial condition. 
\begin{figure}
\centering
\includegraphics[width=0.9\columnwidth]{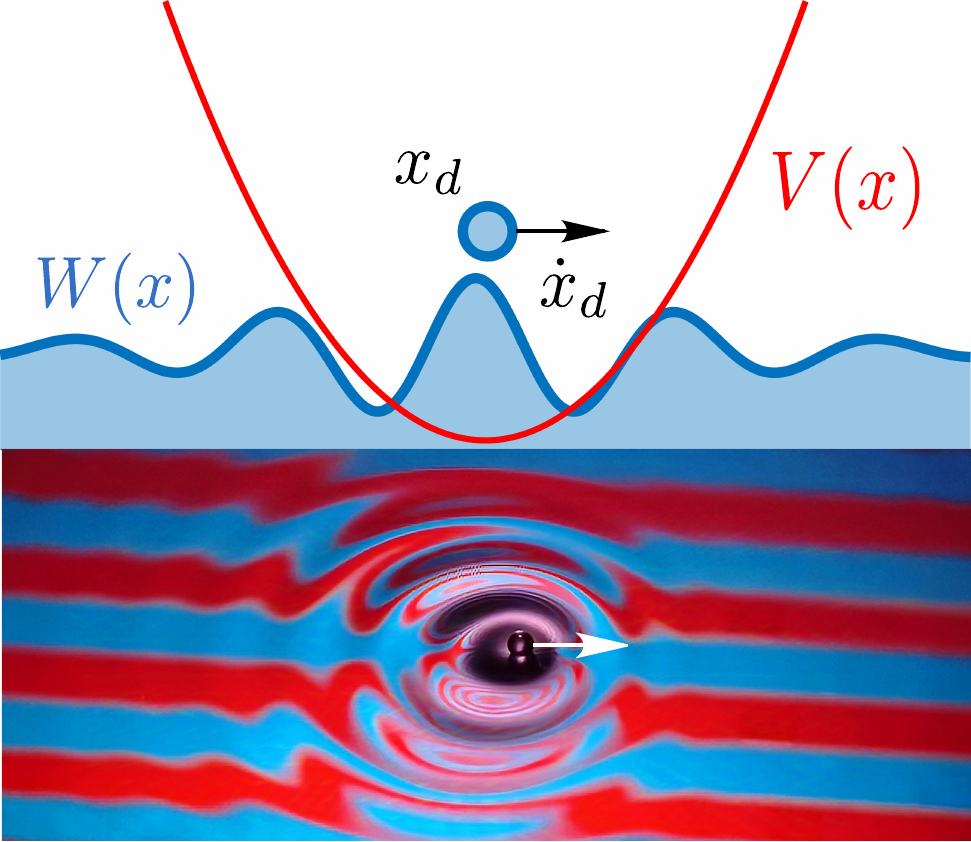}
\caption{A walking droplet immersed in a harmonic external potential $V(x)$ represented by the red curve, and its self-generated wave profile $W(x)$ in blue. When the droplet impacts on the fluid, the slope of the wave field produces a self-excited force. In the bottom figure we show a real drop with its guiding pilot wave.}
\label{Fig:1}
\end{figure}

\section{Generalized pilot-wave hydrodynamics}\label{Sec: DS}

We now apply the previous results to a dynamical system that has recently gained considerable physical interest, namely hydrodynamic wave-particle entities~\citep{Couder2005WalkingDroplets,superwalker}. 


By vertically vibrating an oil bath, a drop of the same oil can bounce and walk on the liquid surface. Each bounce of the droplet locally excites a damped standing wave. The droplet interacts with these self-excited waves on subsequent bounces to propel itself horizontally, giving rise to an active (self-propelled) classical wave-particle entity (WPE). At large vibration amplitudes, droplet-generated waves decay slowly in time. Hence, the motion of the droplet is affected by the history of waves along its trajectory. This gives rise to \textit{path memory} in the system and makes the dynamics non-Markovian. 

Oza \textit{et al.}~\citep{Oza2013} developed a theoretical stroboscopic model to describe the horizontal walking motion of such a WPE. The model averages over the fast vertical periodic bouncing of the droplet and provides a trajectory equation for the slow-walking dynamics in the horizontal plane. We consider a reduction of this model to one horizontal dimension with an added external harmonic potential. As shown in Fig.~\ref{Fig:1}, consider a droplet with position and velocity given by $(x(t), \dot{x}(t))$ in a harmonic potential, which continuously generates standing waves with prescribed spatial structure $W(x)$ that decay exponentially in time. The dynamics of the $1$D WPE follows the non-dimensional integro-differential equation
\begin{align}\label{Eq: dimeless}
   & \ddot{x}+\mu\dot{x} + K x =\\ \nonumber 
    &-\beta\int_{-\infty}^{t-\tau} W'\left(x(t)-x(s)\right)\,\text{e}^{-\frac{(t-s)}{ \text{Me}}}\,\text{d}s. 
\end{align}
where $\mu$ is a dimensionless drag coefficient, $K$ is a dimensionless spring constant, $\tau$ is a cut-off memory time,  $\beta$ is a dimensionless wave-amplitude parameter and $\text{Me}$ is a memory parameter. We refer the reader to Refs.~\citep{LopezValani2025,Oza2013} for details and explicit expressions for these parameters.  Eq.~\eqref{Eq: dimeless} is a horizontal force balance of the WPE with the left-hand side containing an inertial term, an effective dissipation term and a spring force term, respectively. The right-hand side consists of the non-local memory term capturing the cumulative force on the particle from the superposition of the self-generated waves along its path. The memory kernel comprises the function $W(x)$, which represents the wave height. In the stroboscopic model of a walking droplet $W(x)=J_{0}(x)$, thus the wave gradient is $W'(x) = - J_1(x)$, where $J_0(x)$ is the Bessel function of the first kind, of order 0.

We consider the low-memory regime where the particle is only influenced by its most recent wave at time $t-\tau$ (see Ref.~\citep{LopezValani2025} for more details) which gives 
\begin{align}
\label{eq:13_b}
\ddot{x}+\mu\dot{x}+K x=-\epsilon W'(\tau\dot{x}),
\end{align}
where $\epsilon=\beta\,\text{e}^{-\frac{1}{\text{Me}}}$. This results in our low-memory self-excited oscillator equation. We note that the above equation with $K=0$ is of the same structure as the seminal model of a walking droplet at low memory proposed by Protière et al.~\cite{Protiere2006}.

We rescale our system so that $K=1$ and fix $\tau=1$ with any loss of generality (the megastable structure rescales as $\tau$ is increased) and rewrite the dynamical system in Eq.~\eqref{eq:13_b} in phase space, as
\begin{align} \label{eq:14}
\dot{x} & = v, \\ \nonumber
\dot{v} & = - x - \mu v +\epsilon J_{1}(v).
\end{align}

For $\mu=0$, the system described by Eq.~\eqref{eq:14} exhibits an unbounded hierarchy of nested limit cycles. The existence of such a discrete but infinite spectrum of limit cycles in dynamical systems literature is known as megastability. We employ the Krylov-Bogoliubov averaging method to derive analytical expressions for this spectrum. The observed megastable quantization arises from a dynamic equilibrium between the external harmonic force and the self-induced wave-mediated force acting on the particle.

Substituting the ansatz into Eq.~\eqref{eq:14} yields the following system for the amplitude and phase dynamics
\begin{align}
\dot{r} & =  \epsilon J_1 (r \cos \theta) \cos \theta , \label{eq:15_1} \\ 
r \dot{\varphi} & = -\epsilon J_1 (r \cos \theta) \sin \theta.
\label{eq:15_2}
\end{align}

Averaging Eqs.~\eqref{eq:15_1} and \eqref{eq:15_2} over one oscillation period $\theta \in [0,2\pi]$ of the unperturbed system yields $\dot{\varphi}=0$ and
\begin{equation}
\dot{r} = \epsilon J_{0}(r/2) J_{1}(r/2).
\label{eq:16}
\end{equation}

\begin{figure}
\centering
\includegraphics[width=0.9\columnwidth]{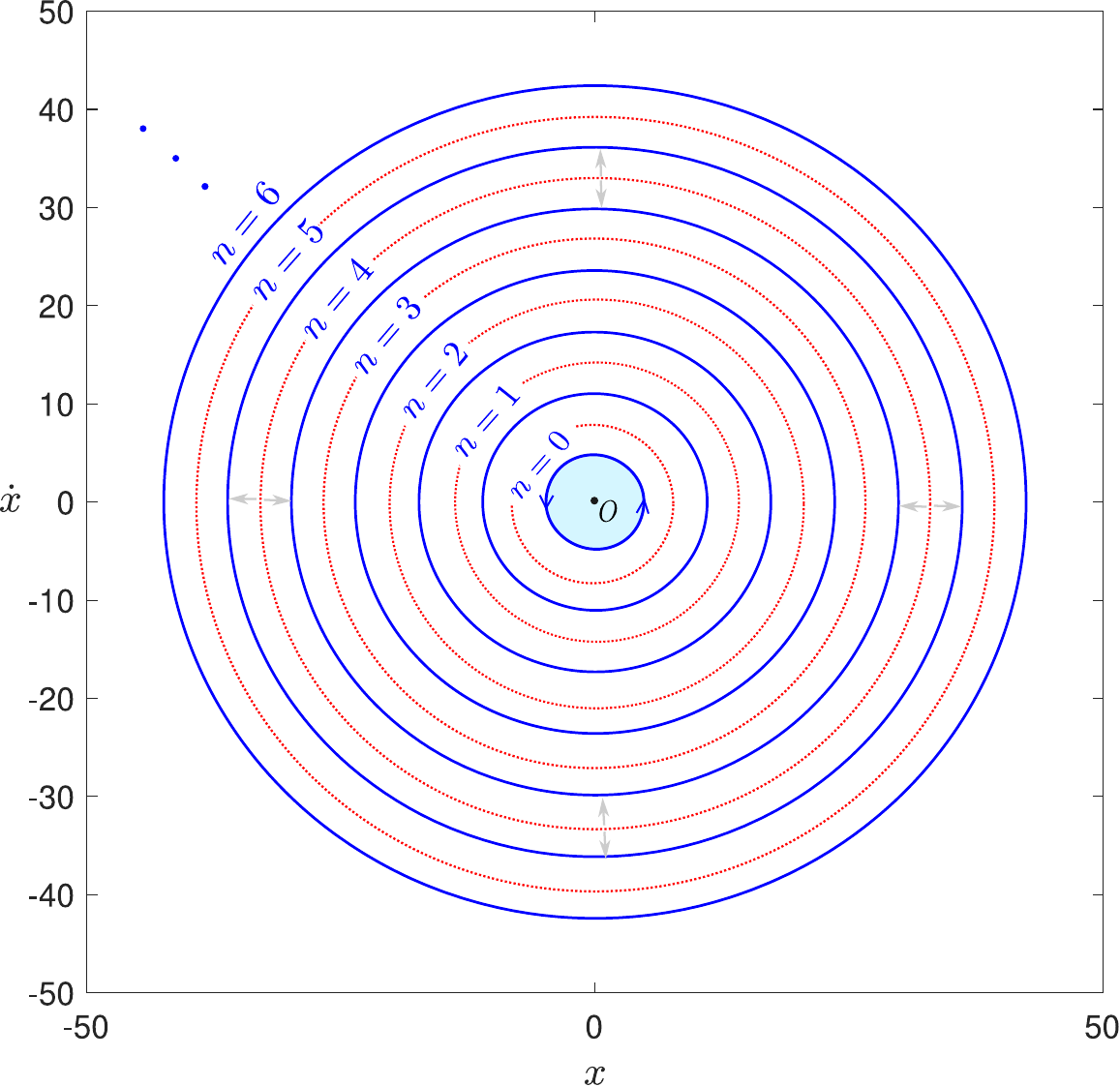}
\caption{A megastable set of quantized orbits, i.e. limit cycles, obtained by numerically solving Eq.~\eqref{eq:14} (dots indicate that the sequence continues). The stable limit cycles (blue), separated by their unstable counterparts (dashed red), which correspond to the basin boundaries (gray arrows). The limit-cycle enclosed area (see light blue for zero-point energy level) is preserved under the flow (blue arrows). The parameter values are fixed to $\epsilon=1/2$, $\mu=0$.}
\label{Fig:2}
\end{figure}

The fixed points of Eq.~\eqref{eq:16} correspond to the infinite set of roots of the product of Bessel functions. For large $r$, this product asymptotically approximates $J_{0}(r/2)J_{1}(r/2) \approx -\pi \cos(r)/r$, implying that the amplitudes $r_n$ of the limit cycles approximate $r_n \approx \pi(n+1/2)$. This structure produces a countably infinite sequence of alternating stable and unstable limit cycles, as shown in Fig.~\ref{Fig:2}. As dissipation $\mu$ increases, an infinite sequence of subcritical saddle-node bifurcations progressively annihilates the larger limit cycles, and only a finite number $N$ of limit cycles persist \cite{LopezValani2025}.

Among the infinite megastable spectrum, the fundamental (lowest-energy) level acts as a self-excited attractor corresponding to zero-point fluctuations. It shares its basin of attraction with the unstable equilibrium at $(x^{*}, v^{*}) = (0,0)$. Higher quantized levels are hidden attractors~\cite{KUZNETSOV20145445}, possessing independent basins of attraction without connection to any fixed point.  In Fig.~\ref{Fig:2}, we numerically calculate limit cycles beyond the first ten levels by initializing trajectories with $(x(0), v(0)) = (A,0)$, for increasing values of $A$ and $\mu=0$, $\epsilon=1/2$. Simulations confirm the megastable structure, with the limit-cycle size exhibiting a linear growth with the principal quantun number $n$.

\begin{figure*}
\centering
\includegraphics[width=1.4\columnwidth]{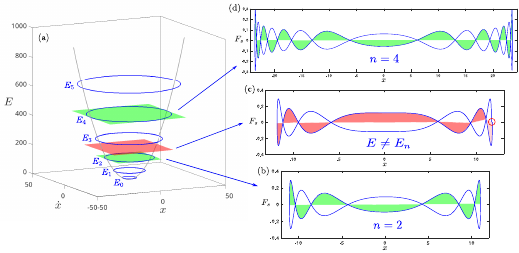}
\caption{(a) The energy shells (blue curves) are shown as a function of phase space coordinates. Some shells are allowed (green planes), while others are forbidden (red plane). (b) The force of the self-excited force is plotted against the spatial coordinate for the third energy level $E_2$, showing that quantized orbits are closed and have zero net work, involving a self-balance of energy. (c) On the contrary, forbidden energy levels involve open curves with tilting nodal points (red circle), implying a mismatch in energy average conservation. (d) Another self-balanced orbit for the fifth energy level $E_4$, with a higher number of self-crossings. The parameter values are fixed to $\epsilon=1/2$, $\mu=0$.}
\label{Fig:3}
\end{figure*}

To characterize the energy spectrum, we derive \emph{quantization relation} that relates the energy and frequency of the oscillator. The oscillator energy is obtained using the Lyapunov function in Eq.~\eqref{eq:2}. Thus, we assign to each stable limit cycle an energy eigenvalue defined by the time-averaged Lyapunov energy function, $E_n = \frac{1}{T} \int_0^T E(x_{n}(t), v_{n}(t))\,dt$, where $(x_{n}(t), v_{n}(t))$ denotes the trajectory of the $n$-th stable limit cycle with period $T$. The corresponding energy levels within the external harmonic potential are shown in Fig.~\ref{Fig:3}. Given that the stable limit cycles are approximately circular, with radius $r_{s,n} \approx \pi (2n + 1/4)$, we estimate the energy spectrum as $E_n \approx r_{s,n}^2/2 = 2\pi^2 (n + 1/8)^2$, resulting in a quadratic dependence of energy on the orbit number $n$. On the other hand, the frequency spectrum of the limit cycles results from the second Krylov-Bogoliubov equation and is given by $\omega_n=1$.

A least squares fit can be used to assess the agreement of this parabolic trend \cite{LopezValani2025}, which is excellent as a consequence of the quasi-harmonic nature of limit cycles (i. e. the smallness of $\epsilon$). 
We now focus on the energy self-balance mechanism. As depicted in Figs.~\ref{Fig:3}(b) and (d), we plot the values of the self-excited force as the trajectory unfolds in physical space for the third $E_2$ and fifth $E_4$ energy levels. As the energy is increased, we observe that the self-force performs more oscillations, yielding a sequence of self-crossing curves. The number of nodal points for the $E_n$ level corresponds to $2n+4$. The total area subtended by the non-conservative force along half of the period (and thus a full period) of the periodic orbit equals zero (the green subtended areas appearing in Fig.~\ref{Fig:3}(b) alternate positive and negative, adding to zero). Consequently, as long as we restrict the dynamics to the limit-cycle orbits, the self-force does no work. However, if we choose an energy value different from the quantized spectrum, an imbalance distorts the average energy conservation. As shown in Fig.~\ref{Fig:3}(c), the work excess (or deficit) yields an open self-crossing curve, which later spirals towards the closest quantized orbit \cite{lopez2024megastable}. The ultimate destination of these unstable orbits depends on the structure of the basins of attraction, as depicted in Fig.~\ref{Fig:2}. Thus, the self-excited force acts as a selective principle, restricting the continuous energy spectrum of the corresponding Hamiltonian system to those closed orbits that balance self-propulsion and dissipation. 

A similar energy analysis has recently been addressed in a study of the energetics of walking droplets \cite{durey2025energetics}, although its crucial link to quantization remains unclear. In that work, the gravitational potential energy of the droplet is considered to be part of the mechanical energy. However, since the Oza model is stroboscopic, this added contribution yields a time-delayed and time-dependent non-Hamiltonian contribution to the energy \cite{durey2025energetics}. In contrast, isolated quantum mechanical systems require as a foundation the symplectic structure provided by Hamiltonian dynamical systems. For this reason, only a two-dimensional correspondence between the quantum mechanical system and pilot-wave hydrodynamics can be strictly established. The role of vertical dynamics is to enable the resonant mechanism that creates the stationary pilot wave, and it is not governed by the Oza equation \cite{Oza2013}. Consequently, we have left out of the Lyapunov energy function the energetic contributions coming from the vertical dynamics, including the energy of the fluid and the external driving force exerted by the shaker, which, apart from generating the self-excited horizontal force, must balance amongst themselves.

\begin{figure*}
\centering
\includegraphics[width=0.8\columnwidth]{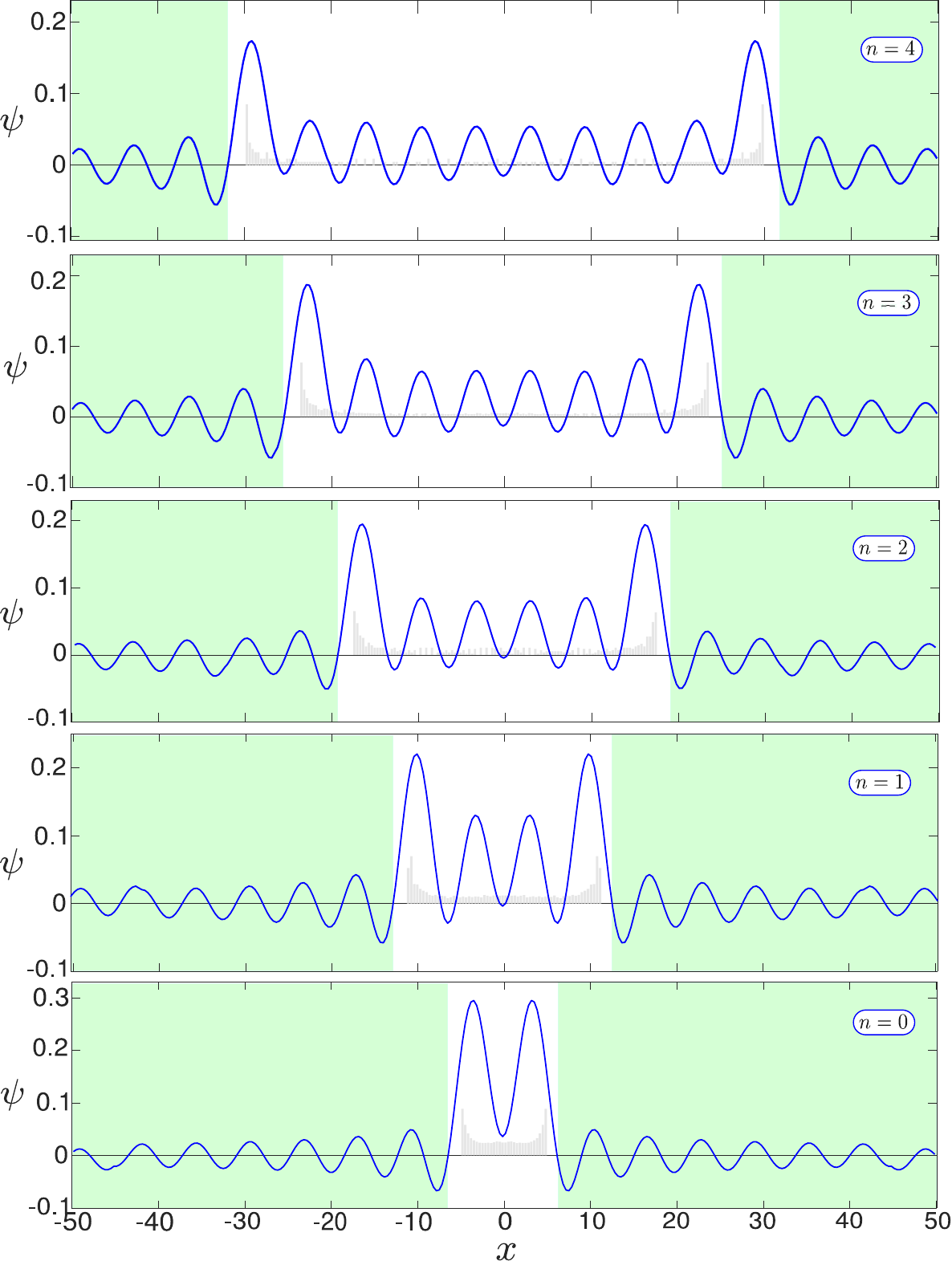}
\caption{The wave eigenfunctions $\psi_n(x)$ for the first five energy levels. As we can see, the number of nodes of the wave function increases with the principal quantum number $n$ following the sequence $4n+4$. The wave function decays rapidly to zero beyond the turning points, even though the probability is no exactly zero, as in the Hamiltonian case. The histograms representing the natural measure $\mu(x)$ are depicted in light gray. The parameter values are fixed to $\epsilon=1/2$, $\mu=0$.}
\label{Fig:4}
\end{figure*}

To conclude this section, we compute the eigenfunctions of our classical quantized system. We take advantage of a theorem that relates the time average of the wave field of the pilot wave $\eta(x,t) = \int_{-\infty}^{t-1} W\left(x(t)-x(s)\right)\,\text{e}^{-\frac{(t-s)}{ \text{Me}}}\,\text{d}s$ and the wave function $\psi_{n}(x)$ of a certain quantized orbit \cite{durey2018}. In the case of periodic motion, this theorem states that the time average of the pilot wave along the limit cycle can be computed from the convolution between the wave profile of a static bouncing droplet $\eta_{B}(x)$ and the natural measure $\mu(x)$ of the droplet's trajectory in the configuration space. The resulting average wave field $\bar \eta_n(x)$ can be identified with the wave function $\psi_n(x)$ of a particular energy level. In the low-memory regime we have
\begin{equation}
\psi_{n}(x) = \int^{\infty}_{-\infty} J_{0}(x-y) \mu_n(y) d y.
\label{eq:17}
\end{equation}
To obtain analytic results, we again consider the asymptotic expression of the Bessel function $J_0(x)=\sqrt{2/\pi x} \cos(x-\pi/4)$, and further approximate it by neglecting the spatial decay. Since the limit cycles are quasi-harmonic, we can use the measure of the harmonic oscillator $\mu(x)=1/\pi \sqrt{r_n^2-x^2}$. This yields the approximate eigenfunctions $\psi_n (x)= \sqrt{2/\pi} H_{0}(r_n)\sin(r_n x+\pi/4)$, where $H_0(x)$ the Hankel function. These normal models are strikingly similar to the wave function of a quantum particle in a box. In Fig.~\ref{Fig:4} we have computed the average wave field $\bar\eta(x)$ using the Bessel function $J_0(x)$ wave form. Again, several differences are appreciated with respect to the quantum-mechanical case. Firstly, only the limit cycle orbits with even numbers are stable, and so is the number of nodes of the wave function, giving a new theorem of oscillations \cite{bransden2000}. Secondly, the Bessel function profile and the measure of the harmonic oscillator imprint two asymmetries, since the heights of the peaks of the wave function are greatest at the extrema (where the oscillator spends larger times) and always biased towards positive real values. Similar results have been obtained in previous studies of Bohr-Sommerfeld quatization in stroboscopic models of walking droplets in the infinite memory regime \cite{PhysRevE.103.053110}. In contrast, here we see it in the low-memory regime.

\section{Discussion and Conclusions}\label{Sec: DS}

It has been shown that self-excited systems can exhibit a countably infinite number of discrete limit cycle orbits, and that these megastable orbits are selected from the non-denumerable set of orbits present of the corresponding continuous conservative physical system ($\epsilon=0$), by a principle of energy balance. We suspect that both the conservation of the phase space area and the average conservation of energy are crucial for classical Hamiltonian mechanics to be applicable on the average along quantized orbits, as originally proposed by Bohr \cite{bohr13} and, more formally, by Wilson and Sommerfeld \cite{Wilson01061915}. 

However, in the present examples, the $\epsilon \ll 1$ limit gives a radius of the quasiharmonic limit cycles that increases linearly with the quantum number \cite{LopezValani2025}. Therefore, the phase space area enclosed by the quantized orbits (and the energy eigenvalues) can be considered to be proportional to the square of the principal quantum number $n^2$. This contrasts to previous studies with walking droplets, where the high-memory limit is exactly matched to the Bohr-Sommerfeld quantization rule \cite{PhysRevE.103.053110}. However, this work does not contain megastable structures, and the different energy levels are obtained by variation of the parameters of the potential. Nevertheless, the spectrum of the harmonic potential can be derived exactly a more sophisticated self-excited force $g(\epsilon,x,y)$ as shown in \cite{ZARMI201721}.

For the class of megastable systems here considered, where $g$ is a pure transcendental function (e.g., trigonometric or Bessel function), a simple renormalization argument allows us to write the parameter $\epsilon$ as small as desired, by rescaling the function $g$. Hence, we confirm that our general results stand when $\epsilon \gg 1$. We have verified this by numerical simulations even for apparently non-renormalizable functions $g(\epsilon,x,y)$, where even or odd polynomials multiply the transcendental function \cite{LopezValani2025}, or even for delayed-differential equations, which can also yield megastable systems \cite{lopez2024megastable}. 

Therefore, the present work suggests a general method to provide quantization rules for more complicated high-dimensional megastable systems, including time-delayed dynamical systems of electrodynamic origin \cite{raju2004electrodynamic, onanelec}, or even explicit spatiotemporal fields \cite{lopezmassive25}, where retarded potentials \cite{lienard1898champ,OSETRIN2024169619} produce wave-mediated interactions between particles. It suffices to demand that the area projections in the \emph{classical} phase space are instantaneously preserved and that the work performed by the self-excited forces equals zero along a closed periodic orbit. Consequently, we conjecture that two general equations describe classical orbit quantization in any self-excited perturbed Hamiltonian system
\begin{equation}
\frac{1}{T}\int^{\vec{x}_i(T)}_{\vec{x}_i(0)} \vec{g} d \vec{x}_i = 0, 
\label{eq:18} 
\end{equation}
where $\vec{g}$ is a generalized self-excited force acting on the system, and $\vec{x}_1= \vec{x}$ is the position vector of the mass center of the particle, while $\vec{x}_2 = \vec{y}$ is the canonical momentum. In the case that the underlying Hamiltonian system is integrable, a canonical transformation allows us to express Eqs.~\eqref{eq:3} and~\eqref{eq:18} in terms of action-angle variables, producing a sequence of $N$ quantum numbers, where $N$ is the dimension of the dynamical system in the configuration space.

Otherwise, when hyperbolic chaotic strange attractors are the asymptotic limit sets of the self-excited system \cite{LOPEZ2023113412}, we can impose a similar condition by recalling that orbits are dense on the attractor \cite{alligood1998chaos}, and taking the limit $T \rightarrow \infty$ in Eq.~\eqref{eq:18} to compute the net work done by the self-excited force and the preservation of phase-space area projections, which in this case would be \emph{conserved on average} and not instantaneously. Simply put, we conjecture that the theorem proposed in Eq.~\eqref{eq:18} can be extended to chaotic attractors by averaging the phase-space area projections over infinite time. This proposition might be proved by recalling that the fractal dimension of a strange attractor \cite{frederickson1983liapunov}, when projected onto a hyperplane of smaller dimension, reduces to the topological dimension of this hyperplane.

In summary, the present work overcomes the drawbacks originally found in the old quantum theory, before the discovery of the Schr\"odinger equation \cite{hawking2011dreams}. Nevertheless, as we have shown, these difficulties can also be circumvented in walking droplet models by considering the complete hydrodynamic representation of the fluid field accompanying the particle, and whose average yields a probability density of the wave-particle entity at the fluid's interface evolving according to a non-local Vlasov-Fokker-Planck equation \cite{durey2018}. Analogously, we might expect that the Schr\"odinger equation describes the dynamical evolution of an averaged electromagnetic field \cite{PhysRev.85.166,onanelec,lopezmassive25}.

\textit{Acknowledgements} We acknowledge Yuanmei Li (Nanjing University) for the photo of the walking droplet appearing in Fig.~\ref{Fig:1}.

 \bibliographystyle{elsarticle-num} 
 \bibliography{MS_harmonic}

\begin{thebibliography}{10}
\expandafter\ifx\csname url\endcsname\relax
  \def\url#1{\texttt{#1}}\fi
\expandafter\ifx\csname urlprefix\endcsname\relax\def\urlprefix{URL }\fi
\expandafter\ifx\csname href\endcsname\relax
  \def\href#1#2{#2} \def\path#1{#1}\fi

\bibitem{Fort_2013}
E.~Fort, Y.~Couder, Trajectory eigenmodes of an orbiting wave source, Europhys.
  Lett. 102~(1) (2013) 16005.

\bibitem{labousse2016b}
M.~Labousse, A.~U. Oza, S.~Perrard, J.~W.~M. Bush, Pilot-wave dynamics in a
  harmonic potential: Quantization and stability of circular orbits, Phys. Rev.
  E 93 (2016) 033122.

\bibitem{Perrard2014a}
S.~Perrard, M.~Labousse, M.~Miskin, E.~Fort, Y.~Couder, Self-organization into
  quantized eigenstates of a classical wave-driven particle, Nat. Commun. 5
  (2014) 3219.

\bibitem{Perrard2014b}
S.~Perrard, M.~Labousse, E.~Fort, Y.~Couder, Chaos driven by interfering
  memory, Phys. Rev. Lett. 113 (2014) 104101.

\bibitem{Tambasco2016}
L.~D. Tambasco, D.~M. Harris, A.~U. Oza, R.~R. Rosales, J.~W.~M. Bush, The
  onset of chaos in orbital pilot-wave dynamics, Chaos 26~(10) (2016) 103107.

\bibitem{Tambascoorbit}
L.~D. Tambasco, J.~W.~M. Bush, Exploring orbital dynamics and trapping with a
  generalized pilot-wave framework, Chaos 28~(9) (2018) 096115.

\bibitem{Labousse_2014}
M.~Labousse, S.~Perrard, Y.~Couder, E.~Fort, Build-up of macroscopic
  eigenstates in a memory-based constrained system, New J. Phys. 16~(11) (2014)
  113027.

\bibitem{labousseharmonic}
M.~Labousse, A.~U. Oza, S.~Perrard, J.~W.~M. Bush, Pilot-wave dynamics in a
  harmonic potential: Quantization and stability of circular orbits, Phys. Rev.
  E 93 (2016) 033122.

\bibitem{Kurianskiharmonic}
K.~M. Kurianski, A.~U. Oza, J.~W.~M. Bush, Simulations of pilot-wave dynamics
  in a simple harmonic potential, Phys. Rev. Fluids 2 (2017) 113602.

\bibitem{durey2018}
M.~Durey, P.~A. Milewski, J.~W.~M. Bush, Dynamics, emergent statistics, and the
  mean-pilot-wave potential of walking droplets, Chaos 28~(9) (2018) 096108.

\bibitem{Perrard2018}
S.~Perrard, M.~Labousse, Transition to chaos in wave memory dynamics in a
  harmonic well: Deterministic and noise-driven behavior, Chaos 28~(9) (2018)
  096109.

\bibitem{Blits24}
A.~M. Blitstein, R.~R. Rosales, P.~J. S\'aenz, Minimal quantization model in
  pilot-wave hydrodynamics, Phys. Rev. Lett. 132 (2024) 104003.

\bibitem{LopezValani2025}
A.~G. L\'opez, R.~N. Valani, Megastable quantization in generalized pilot-wave
  hydrodynamics, Phys. Rev. E 111 (2025) L022201.

\bibitem{Sprott2017}
J.~C. Sprott, S.~Jafari, A.~J.~M. Khalaf, T.~Kapitaniak, Megastability:
  Coexistence of a countable infinity of nested attractors in a
  periodically-forced oscillator with spatially-periodic damping, Eur. Phys. J.
  Spec. Top. 226~(9) (2017) 1979--1985.

\bibitem{jenkins}
A.~Jenkins, Self-oscillation, Phys. Rep. 525~(2) (2013) 167--222.

\bibitem{lopezte}
A.~G. López, F.~Benito, J.~Sabuco, A.~Delgado-Bonal, The thermodynamic
  efficiency of the {Lorenz} system, Chaos, Solit. Fractals 172 (2023) 113521.

\bibitem{bohr13}
N.~Bohr, I. {On} the constitution of atoms and molecules, The London,
  Edinburgh, and Dublin Philosophical Magazine and Journal of Science 26~(151)
  (1913) 1--25.

\bibitem{Wilson01061915}
W.~Wilson, The quantum-theory of radiation and line spectra, The London,
  Edinburgh, and Dublin Philosophical Magazine and Journal of Science 29~(174)
  (1915) 795--802.

\bibitem{VALANI2024115253}
R.~N. Valani, Álvaro G.~López, Quantum-like behavior of an active particle in
  a double-well potential, Chaos, Solitons \& Fractals 186 (2024) 115253.

\bibitem{poincare1908}
H.~Poincaré, Sur la télégraphie sans fil, La Lumière Électrique 4~(II)
  (1908) 387--393.

\bibitem{sanjuan1998lienard}
M.~A. Sanju{\'a}n, Li{\'e}nard systems, limit cycles, and {M}elnikov theory,
  Physical Review E 57~(1) (1998) 340.

\bibitem{Davidow2017}
M.~Davidow, B.~Shayak, R.~H. Rand, Analysis of a remarkable singularity in a
  nonlinear {DDE}, Nonlinear Dyn. 90~(1) (2017) 317--323.

\bibitem{guckenheimer2013nonlinear}
J.~Guckenheimer, P.~Holmes, Nonlinear oscillations, dynamical systems, and
  bifurcations of vector fields, Vol.~42, Springer Science \& Business Media,
  2013, pp. 184---204.

\bibitem{sommerfeld1923}
A.~Sommerfeld, Atomic Structure and Spectral Lines, Dutton, New York, 1923, pp.
  193---203.

\bibitem{han2009hopf}
M.~Han, J.~Yang, P.~Yu, Hopf bifurcations for near-{H}amiltonian systems,
  International Journal of Bifurcation and Chaos 19~(12) (2009) 4117--4130.

\bibitem{krylov1950}
N.~M. Krylov, N.~N. Bogoliubov, Introduction to non-linear mechanics, no.~11,
  Princeton university press, 1950.

\bibitem{pikovsky1985universal}
A.~Pikovsky, M.~Rosenblum, S.~KJ, Synchronization: A universal concept in
  nonlinear sciences, Cambridge University Press, Princeton, 1985, pp. 32---35.

\bibitem{papatryfonos2024static}
K.~Papatryfonos, L.~Vervoort, A.~Nachbin, M.~Labousse, J.~W. Bush, Static bell
  test in pilot-wave hydrodynamics, Physical Review Fluids 9~(8) (2024) 084001.

\bibitem{Couder2005WalkingDroplets}
Y.~Couder, S.~Proti{\`{e}}re, E.~Fort, A.~Boudaoud, Dynamical phenomena:
  Walking and orbiting droplets, Nature 437~(7056) (2005) 208--208.

\bibitem{superwalker}
R.~N. Valani, A.~C. Slim, T.~Simula, Superwalking droplets, Phys. Rev. Lett.
  123 (2019) 024503.

\bibitem{Oza2013}
A.~U. Oza, R.~R. Rosales, J.~W.~M. Bush, A trajectory equation for walking
  droplets: hydrodynamic pilot-wave theory, J. Fluid Mech. 737 (2013) 552--570.

\bibitem{Protiere2006}
S.~Proti{\`e}re, A.~Boudaoud, Y.~Couder, Particle--wave association on a fluid
  interface, J. Fluid Mech. 554 (2006) 85--108.

\bibitem{KUZNETSOV20145445}
N.~Kuznetsov, G.~Leonov, Hidden attractors in dynamical systems: systems with
  no equilibria, multistability and coexisting attractors, IFAC Proceedings
  Volumes 47~(3) (2014) 5445--5454.

\bibitem{lopez2024megastable}
{\'A}.~G. L{\'o}pez, R.~N. Valani, Driven transitions between megastable
  quantized orbits, arXiv preprint arXiv:2406.03906v4 (2024).

\bibitem{durey2025energetics}
M.~Durey, J.~W. Bush, The energetics of pilot-wave hydrodynamics, Journal of
  Fluid Mechanics 1009 (2025) A4.

\bibitem{bransden2000}
B.~H. Bransden, C.~J. Joachain, Quantum Mechanics, 2nd Edition, Pearson
  Education, Harlow, England, 2000, pp. 118--119.

\bibitem{PhysRevE.103.053110}
J.~Montes, F.~Revuelta, F.~Borondo, Bohr-{S}ommerfeld-like quantization in the
  theory of walking droplets, Phys. Rev. E 103 (2021) 053110.

\bibitem{ZARMI201721}
Y.~Zarmi, A classical limit-cycle system that mimics the quantum-mechanical
  harmonic oscillator, Phys. D: Nonlinear Phenom. 359 (2017) 21--28.

\bibitem{raju2004electrodynamic}
C.~K. Raju, The electrodynamic 2-body problem and the origin of quantum
  mechanics, Foundations of Physics 34~(6) (2004) 937--963.

\bibitem{onanelec}
A.~G. L\'{o}pez, On an electrodynamic origin of quantum fluctuations, Nonlinear
  Dyn. 102 (2020) 621--634.

\bibitem{lopezmassive25}
A.~G. Lopez, Massive wave solutions to the {E}instein-{M}axwell equations,
  Preprints (2025).

\bibitem{lienard1898champ}
A.-M. Liénard, Champ électrique et magnétique produit par une charge
  concentrée en un point et animée d'un mouvement quelconque, L'Éclairage
  électrique 16 (1898) 5--21, 53--59, 106--112.

\bibitem{OSETRIN2024169619}
K.~E. Osetrin, V.~Y. Epp, S.~V. Chervon, Propagation of light and retarded time
  of radiation in a strong gravitational wave, Annals of Physics 462 (2024)
  169619.

\bibitem{LOPEZ2023113412}
A.~G. L\'{o}pez, Orbit quantization in a retarded harmonic oscillator, Chaos
  Solit. Fractals 170 (2023) 113412.

\bibitem{alligood1998chaos}
K.~T. Alligood, T.~D. Sauer, J.~A. Yorke, D.~Chillingworth, Chaos: an
  introduction to dynamical systems, Vol.~40, Philadelphia, Society for
  Industrial and Applied Mathematics., 1998, pp. 114--124.

\bibitem{frederickson1983liapunov}
P.~Frederickson, J.~L. Kaplan, E.~D. Yorke, J.~A. Yorke, The {L}iapunov
  dimension of strange attractors, Journal of differential equations 49~(2)
  (1983) 185--207.

\bibitem{hawking2011dreams}
S.~Hawking, The dreams that stuff is made of: The most astounding papers of
  quantum physics—and how they shook the scientific world, Running Press,
  Philadelphia, PA, 2011, pp. 251---386.

\bibitem{PhysRev.85.166}
D.~Bohm, A suggested interpretation of the quantum theory in terms of "hidden"
  variables. i, Phys. Rev. 85 (1952) 166--179.

\end{thebibliography}




\biboptions{sort&compress}
\end{document}